\documentclass[12pt]{article}
\usepackage{latexsym}
\usepackage{amsmath}
\usepackage{amssymb}
\usepackage{amssymb}
\catcode`\@=11
\newcount\hour
\newcount\minute
\newtoks\amorpm \hour=\time\divide\hour by 60\minute
=\time{\multiply\hour by 60 \global\advance\minute by-\hour}
\edef\standardtime{{\ifnum\hour<12 \global\amorpm={am}%
        \else\global\amorpm={pm}\advance\hour by-12 \fi
        \ifnum\hour=0 \hour=12 \fi
        \number\hour:\ifnum\minute<10
        0\fi\number\minute\the\amorpm}}
\edef\militarytime{\number\hour:\ifnum\minute<10
0\fi\number\minute}
\def\draftlabel#1{{\@bsphack\if@filesw {\let\thepage\relax
   \xdef\@gtempa{\write\@auxout{\string
      \newlabel{#1}{{\@currentlabel}{\thepage}}}}}\@gtempa
   \if@nobreak \ifvmode\nobreak\fi\fi\fi\@esphack}
        \gdef\@eqnlabel{#1}}
\def\@eqnlabel{}
\def\@vacuum{}
\def\marginnote#1{}
\def\draftmarginnote#1{\marginpar{\raggedright\scriptsize\tt#1}}
\def\draft{
        \pagestyle{plain}
        \overfullrule=2pt
        \oddsidemargin -.1truein
        \def\@oddhead{\sl \phantom{\today\quad\militarytime} \hfil
        \smash{\Large\sl DRAFT} \hfil \today\quad\militarytime}
        \let\@evenhead\@oddhead
        \let\label=\draftlabel
        \let\marginnote=\draftmarginnote
        \def\ps@empty{\let\@mkboth\@gobbletwo
        \def\@oddfoot{\hfil \smash{\Large\sl DRAFT} \hfil}
        \let\@evenfoot\@oddhead}
        \def\@eqnnum{(\theequation)\rlap{\kern\marginparsep\tt\@eqnlabel}%
        \global\let\@eqnlabel\@vacuum}  }

\renewcommand{\theequation}{\thesection.\arabic{equation}}
\renewcommand{\thefootnote}{\fnsymbol{footnote}}
\newcommand{\newsection}{    
\setcounter{equation}{0}\section}
\def\appendix#1{\addtocounter{section}{1}\setcounter{equation}{0}
\renewcommand{\thesection}{\Alph{section}}
\section*{Appendix \thesection\protect\indent \parbox[t]{11.15cm}{#1}}
\addcontentsline{toc}{section}{Appendix \thesection\ \ \ #1}}

\def \bi{\bibitem}
\def \la {\label}

\def \b {\beta}

\def\be{\begin{equation}}
\def\ee{\end{equation}}

\catcode`\@=11

\def\bea{\begin{eqnarray}}
\def\eea{\end{eqnarray}}
\def\beann{\begin{eqnarray*}}
\def\eeann{\end{eqnarray*}}
\def\beq{\begin{equation}}
\def\eeq{\end{equation}}
\def\ba{\begin{array}}
\def\ea{\end{array}}
\def\ben{\begin{enumerate}}
\def\een{\end{enumerate}}
 \def \l {\lambda}

 \def \la {\label}
 \def\be{\begin{equation}}
\def\ee{\end{equation}}

\def \la {\label}

\font\mybb=msbm10 at 11pt

\def\bb#1{\hbox{\mybb#1}}

\def\bZ {\bb{Z}}
\def\bR {\bb{R}}

\def\bH {\bb{H}}

\def\e  {\epsilon}

\def \ee {\epsilon}

\def \bi{\bibitem}
\def\a{\alpha }

\def \l {\lambda}

\def \b {\beta}

\def\mt{\eta}

\def\pp {{=\mkern-9mu|}}

\def\be{\begin{equation}}
\def\ee{\end{equation}}

\def \bi {\bibitem}
\def \la{\label}

\begin{document}
\date{November 2002}
\begin{titlepage}
\begin{center}
\vspace{5.0cm}

\vspace{3.0cm} {\Large \bf New half supersymmetric solutions of the heterotic string }
\\
[.2cm]


{}\vspace{2.0cm}
 {\large
 G.~Papadopoulos
 }



{}

\vspace{1.0cm}
Department of Mathematics\\
King's College London\\
Strand\\
London WC2R 2LS, UK\\


\end{center}
{}
\vskip 3.0 cm
\begin{abstract}
We describe all supersymmetric solutions of the heterotic string which preserve 8 supersymmetries and show that
are distinguished by the holonomy, ${\rm hol}(\hat\nabla)$,
 of the connection, $\hat\nabla$, with skew-symmetric torsion.
 The ${\rm hol}(\hat\nabla) \subseteq SU(2)$
 solutions
 are principal bundles over a 4-dimensional hyper-K\"ahler manifold equipped with a anti-self-dual connection and
 fibre group $G$
 which has Lie algebra, ${\mathfrak Lie} (G)=\bR^{5,1}$,  $\mathfrak{sl}(2,\bR)\oplus \mathfrak{su}(2)$ or $\mathfrak{cw}_6$.
 Some of the solutions have the interpretation as 5-branes
wrapped on $G$ with transverse space any hyper-K\"ahler 4-dimensional manifold. We  construct new solutions
for  ${\mathfrak Lie} (G)=\mathfrak{sl}(2,\bR)\oplus \mathfrak{su}(2)$  and show that are characterized by 3 integers
and have continuous moduli. There is also a smooth family in this class
with one asymptotic region and the dilaton is bounded everywhere on the spacetime.

We also demonstrate that the worldvolume theory of the backgrounds with holonomy $SU(2)$ can be understood in terms
of gauged WZW models for which the gauge fields are composite.

The  ${\rm hol}(\hat\nabla) \subseteq\bR^8$ solutions are superpositions of   fundamental strings and  pp-waves  in flat space,
which may also include a null rotation. The ${\rm hol}(\hat\nabla)=\{1\}$ heterotic string backgrounds which preserve 8 supersymmetries
are Lorentzian group manifolds.

\end{abstract}
\end{titlepage}
\newpage
\setcounter{page}{1}
\renewcommand{\thefootnote}{\arabic{footnote}}
\setcounter{footnote}{0}

\setcounter{tocdepth}{1}
\tableofcontents

\setcounter{section}{0}
\setcounter{subsection}{0}
\newpage
\newsection{Introduction}

It is well known for sometime that branes have a low energy description in terms of half supersymmetric
solutions of  10- and 11-dimensional supergravity theories. As such  they have been instrumental
in understanding the superstring dualities and in formulating the proposal for  M-theory, see
\cite{townsend1, townsend2}.
In type II theories, apart from the various branes  \cite{GibbonsMaeda}-\cite{gueven},
half supersymmetric solutions
include  brane bound states, see eg
\cite{dyon}, and   backgrounds  which have applications in AdS/CFT correspondence, see eg  \cite{LLM, dhoker}. Other
half supersymmetric
supergravity solutions are the vacua of
string theory and M-theory compactifications like those on $K_3$. In type I and heterotic theories, it
is known that half supersymmetric
solutions include the fundamental string \cite{rr}, the pp-wave, and the 5-brane \cite{callan}.
Despite these developments our understanding of half supersymmetric solutions of 10- and 11-dimensional
supergravity theories is rather limited.
In particular, it is not known what kind of geometry the half supersymmetric solutions have or a
systematic way to describe them.

In this paper, we  construct all the half supersymmetric solutions of the heterotic\footnote{The
Killing spinor equations of type I supergravity are identical to those of the heterotic string. Consequently,
the half supersymmetric solutions of type I supergravity will be constructed as well. }
 supergravity.
This is based on the results of \cite{het1, het2}, where the Killing spinor equations of heterotic
supergravity has been solved
in all cases using the spinorial geometry technique of \cite{ggp}. We find that there are three
classes of solutions distinguished
by the holonomy, ${\rm hol}(\hat\nabla)$, of the connection, $\hat\nabla$, with skew-symmetric torsion, $H$. These
are ${\rm hol}(\hat\nabla)\subseteq SU(2)$, ${\rm hol}(\hat\nabla)\subseteq \bR^8$ and ${\rm hol}(\hat\nabla)=\{1\}$.

The solutions
of the first class,  ${\rm hol}(\hat\nabla)\subseteq SU(2)$,
 can be  described in terms of the data of a principal bundle. In particular, we show that
  the spacetime $M$ is a principal
bundle, $M=P(G,B_{\rm hk};\pi)$, equipped with a
principal bundle connection $\lambda$. In addition, the base space $B_{\rm hk}$ is a 4-dimensional
{\it hyper-K\"ahler} manifold, and the curvature of $\lambda$ is (1,1)
with respect to all three complex structures, ie $\lambda$ is anti-self-dual. The fibre (gauge) group
$G$ is a 6-dimensional Lorentzian,
{\it self-dual}\footnote{Self-duality in this context means that the 3-form associated with
the structure constants of the Lorentzian Lie algebra, $\mathfrak{Lie}(G)$, of $G$ is self-dual.}
Lie group. The associated self-dual Lie algebras have been classified and have been shown to be isomorphic to
 $\bR^{5,1}$, $\mathfrak{sl}(2, \bR)\oplus
\mathfrak{su}(2)$ or $\mathfrak{cw}_6$ \cite{josela}. In addition, the gaugino Killing spinor equation implies that
 the gauge connection $A$ of the heterotic string is  also a anti-self-dual connection  over $B_{hk}$.
 The Killing spinor equations determine the ${\rm hol}(\hat\nabla)\subseteq SU(2)$ solutions up to an arbitrary
 function $h$ of spacetime. This in turn can be specified by either solving the field equations or the
 the (anomalous) Bianchi identity of $H$ (\ref{fineqn}).
 Therefore the data needed to construct all half supersymmetric solutions
in this class of the heterotic string up and including 2-loops in the sigma model perturbation theory are

\begin{itemize}

\item a hyper-K\"ahler 4-dimensional manifold $B_{\rm hk}$,

\item a anti-self-dual connection $\lambda$ on $B_{\rm hk}$ with gauge group $G$ such that
$\mathfrak{Lie}(G)$ is the  Lorentzian self-dual Lie algebra $\bR^{5,1}$, $\mathfrak{sl}(2, \bR)\oplus
\mathfrak{su}(2)$ or $\mathfrak{cw}_6$,

\item an anti-self-dual connection $A$ on $B_{hk}$,  and

\item a solution of (\ref{fineqn}) for the function $h$.
\end{itemize}
\
Given these data, the solutions can be written\footnote{From now on, the subscript hk indicates that
the associated operation or space is taken with respect to the hyper-K\"ahler metric.} as
\bea
ds^2&=&\mt_{ab} \lambda^a \lambda^b+h\, ds_{\rm hk}^2~,~~~~e^{2\Phi}=h~,
\cr
H&=&{1\over3} \mt_{ab} \lambda^a\wedge d\lambda^b+{2\over3} \mt_{ab} \lambda^a\wedge {\cal F}^b-\star_{\rm hk} dh~,
\la{xxy}
\eea
where ${\cal F}$ is the curvature of $\l$.

We mostly explore explicit solutions for the class of backgrounds for which the contribution
from the anomaly cancelation
mechanism to the Bianchi identity of the 3-form field strength, $H$,  vanishes,
and so $H$ is closed $dH=0$. Although our formalism applies to the case that $dH\not=0$ as well.

The most straightforward case to consider is that for which the principal bundle
connection $\lambda$ is trivial. Then the solutions up to discrete identifications are
\bea
G\times B_{\rm hk}~,~~~5_{B_{hk}}[G]~;~~~G=\bR^{5,1}~,~AdS_3\times S^3~,~CW_6~,
\eea
where $B_{hk}$ is any 4-dimensional hyper-K\"ahler manifold and $5_{B_{hk}}[G]$ denotes the
5-brane solution with transverse space $B_{hk}$ and worldvolume geometry $G$.
For the 5-brane solution of  \cite{callan}, $B_{hk}=\bR^4$ and $G=\bR^{5,1}$.
In the $AdS_3\times S^3\times B_{\rm hk}$  and $5_{B_{hk}}[AdS_3\times S^3]$ solutions,
the radius of $AdS_3$ is equal to that of $S^3$ because
of the self-duality condition. The solutions $CW_6\times B_{\rm hk}$ and $5_{B_{hk}}[G]$, $G\not=\bR^{5,1}$,  are new.

New families of solutions can be constructed whenever $\lambda$ has non-vanishing curvature. Such
solutions have the interpretation of wrapped 5-branes on $G$ with transverse space
any hyper-K\"ahler 4-dimensional manifold $B_{hk}$, where $G=\bR^{5,1}$, $AdS_3\times S^3$ or $CW_6$
up to discrete identifications. In particular, we explicitly construct the solution for $G=AdS_3\times S^3$ and
 $B_{\rm hk}=\bR^4$ in the case that only the $S^3$ is gauged. The background is as in (\ref{xxy}) and $h$
 is given in (\ref{multisol5}).
 These solutions are characterized by three integers $(k,\nu,p)$, where  $k$ is the flux
of
the 3-form in $S^3$, $\nu$ is the instanton number
and $p$ the asymptotic 5-brane charge. They also depend on at least $8\nu-3$ continuous parameters, the moduli
of $SU(2)$ instantons. In addition there is a family of solutions
within this class, with $h$  given in (\ref{xxx}) and $\nu=1$,   which are smooth and the dilaton is bounded over
the whole of spacetime.
In this case $p=4$.
For this family, the spacetime metric has an $AdS_3\times S^3\times \bR^4$
asymptotic region and there is no throat
at the position of the 5-branes. We propose an M-theoretic interpretation of the $\nu=1$
solution as the near M2-brane geometry of a
M2- with -M5 brane intersection on a self-dual string. In particular, we identify the near-horizon geometry
of the M2-brane
as it emerges
from the prescribed data. It is also likely that the more general solution based on  $\nu$ instantons, with $h$
given in (\ref{multisol}), is also smooth and the dilaton bounded over the whole of  spacetime. This solution has $p=4\nu$
and  similar asymptotic properties to the solution based on a single instanton.

The worldvolume theory of the ${\rm hol}(\hat\nabla)\subseteq SU(2)$ class of solutions
can be related to gauged WZW models. In particular,
the principal bundle connection can be thought of as the gauge field which arises from the
gauging of the right action of
a $\bR^{5,1}$, $\mathfrak{sl}(2, \bR)\oplus
\mathfrak{su}(2)$ or $\mathfrak{cw}_6$ WZW model. Such a gauging is anomalous but in this
case the gauge fields are composite
and depend on the scalars of the base space $B_{\rm hk}$. As a result, the anomaly is cancelled
because there is an additional contribution from $B_{\rm hk}$.

The worldsheet supersymmetry of
the resulting theory depends on the particular solution.
If the gauge connection is identified with the spacetime one, ie the contribution of
the anomaly vanishes and so $dH=0$,
the minimal amount of worldvolume supersymmetry is (1,1)
which in some sectors can enhance up to (4,4). All the solutions are
finite up and including 2 loops. In addition an indirect argument, based on
(4,1) worldvolume supersymmetry \cite{howepap1}, suggests that most of, and perhaps all, such
holonomy $SU(2)$ backgrounds are ultraviolet finite to all orders in perturbation theory.
In addition, we comment on the integrability of strings propagating in such backgrounds.

Alternatively, if $dH\not=0$, ie there is an anomalous contribution in the Bianchi identity of $H$, then the minimal
 worldvolume
supersymmetry of the solutions is (1,0) which in some sectors can enhance to (4,0). Again an indirect argument,
based on (4,0) worldvolume supersymmetry \cite{howepap1}, suggests
that some of  these backgrounds are again ultraviolet finite to all orders in perturbation theory.
However, the couplings
may receive finite local counterterm corrections which arise from changing the renormalization scheme from a manifestly
(1,0) to a (4,0) supersymmetric one \cite{howepap2}.

The ${\rm hol}(\hat\nabla)\subseteq \bR^8$ class of solutions include the fundamental string  of \cite{rr} and the pp-wave
solutions in flat space. We show that the most general solutions in this class are superpositions
of the fundamental string, the pp-wave solutions together with a null rotation.
Because of the chiral nature of these models,
$dH=0$, ie the anomalous contribution to the Bianchi identity of $H$ vanishes. The worldvolume theory
of generic such backgrounds exhibits (1,0) supersymmetry. The ultraviolet properties
of the worldvolume theories of these backgrounds have been investigated in \cite{tseytlin1} and it has been
found that some of them are not finite
in the context of heterotic string.

The solutions of the third class of backgrounds that preserve 8 supersymmetries, ${\rm hol}(\hat\nabla)=\{1\}$,
 are WZW models. The spacetime is a Lorentzian  group manifold. The gravitino Killing spinor equation admits
 16 $\hat\nabla$-parallel spinors but the dilatino and the gaugino Killing spinor equations preserve only half of the parallel spinors.
 The spacetime is a product, up to the appropriate dimension,  of one of the Lorentzian groups $SL(2,\bR)$, $\bR$ and $CW_{2n}$
 with the Riemannian groups $\bR$, $SU(2)$ and $SU(3)$. The complete list can be found in table 4 of \cite{josek}.

This paper is organized as follows. In section two, we state the Killing spinor equations of heterotic string. In section
three, we solve the Killing spinor equations for the ${\rm hol}(\hat\nabla)\subseteq SU(2)$ class
of backgrounds. In sections four and five, we give examples of new solutions and explore their properties. In section
six, we give an M-theoretic interpretation for the ${\rm hol}(\hat\nabla)\subseteq SU(2)$ class
of backgrounds. In section seven, we investigate the worldvolume theories associated with the
${\rm hol}(\hat\nabla)\subseteq SU(2)$ backgrounds. In section eight, we give the solutions of
${\rm hol}(\hat\nabla)\subseteq \bR^8$ class of backgrounds, and in section nine we present our conclusions.

\newsection{Killing spinor equations}

The gravitino, dilatino and gaugino Killing spinor equations of the  heterotic supergravity (and type I)  are
\bea
\hat\nabla_A \epsilon&=&0+{\cal O}(\a'^2)~,~~~
\cr
(\Gamma^A \partial_A\Phi-{1\over12} H_{ABC} \Gamma^{ABC})\e&=&0+{\cal O}(\a'^2)
\cr
F_{AB} \Gamma^{AB}\e&=&0+{\cal O}(\a'^2)
\la{kse}
\eea
where $A,B,C=0,\dots 9$ are frame indices, $\Phi$ is the dilaton, $H$ is the NS$\otimes$NS three-form field strength, $F$ is the gauge field strength and
\bea
\hat\nabla_B Y^A=\nabla_B Y^A+{1\over2} H^A{}_{BC} Y^C~,
\eea
is a metric connection with torsion $H$. These Killing spinor equations  are valid  \cite{berg} up and including 2-loops
in the sigma model
perturbation theory \cite{ht}. As it has already been indicated they are expected to receive $\a'$ corrections from higher loops.
Taking into account the 1-loop anomaly cancelation condition, $H$ is not closed but instead
\bea
dH=-{\a'\over4} {\rm tr} (\check R^2-F^2)+{\cal O}(\a'^2)~,
\la{anb}
\eea
where $\check R$ is the curvature of the metric connection $\check \nabla$ which has torsion $-H$. We follow the conventions of
\cite{tsimpis}.

The  Killing spinor equations (\ref{kse}) have been solved in complete generality in \cite{het1, het2} and all the conditions on the
geometry of the spacetime required for the existence of Killing spinors have been found. Here we shall investigate the special case
of backgrounds with strictly 8 supersymmetries. A consequence of the results of \cite{het1, het2} is that such backgrounds
are completely characterized by the holonomy of $\hat\nabla$ connection which can be either $SU(2)$ or $\bR^8$, ie
${\rm hol}(\hat\nabla)\subseteq SU(2), \bR^8$. In what follows, we shall solve the field equations for these
backgrounds and we shall obtain a complete description of the solutions.

\newsection{Holonomy $SU(2)$ solutions}

\subsection{Geometry}

The backgrounds with 8 supersymmetries and ${\rm hol}(\hat\nabla)\subseteq SU(2)$ \cite{het1} admit an orthonormal frame $e^A=(e^a, e^i)$,
 $a=0,5,1,6,2,7$, $i=3,8,4,9$, such
that the six  1-forms $e^a$
and three  2-forms $\omega_r$,  $r=1,2,3$
\bea
e^a~,~~~\omega_r={1\over2}(\omega_r)_{ij}\, e^i\wedge e^j~,~~~~
\eea
are $\hat\nabla$-parallel, where $(\omega_r)_{ij}=\delta_{ik} (I_r)^k{}_j$ and the three endomorphism
$I_r$ satisfy the algebra of imaginary unit quaternions,  $I_r I_s=-\delta_{rs} {\bf 1}_{4\times 4}+\e_{rst} I_t$.
 In addition,
 $e^0$  is chosen to be time-like and the other
five are space-like.

All the conditions that arise from the Killing spinor equations can be expressed as restrictions on the forms
$(e^a, \omega_r)$. Introducing a Hermitian basis with respect to $I_1$, ie $e^i=(e^n, e^{\bar n})$, $n=3,4$,
and $\omega_1=-i \sum_n e^n\wedge e^{\bar n}$,
the conditions on the geometry imposed by gravitino and dilatino Killing spinor equations can be expressed as
\bea
&&\hat\nabla_M e^a_N=0~,~~~\hat\nabla_M \omega_r=0~,~~
\la{dn=8su2g}
\eea
\bea
&&[e_a, e_b]_i=0~,~~~H_{a_1a_2a_3}+{1\over3!} \epsilon_{a_1a_2a_3}{}^{b_1b_2b_3} H_{b_1b_2b_3}=0~,
\cr
&&
(de^a)_n{}^n=0~,~~~(de^a)_{mn}=0~,
\cr
&&2\partial_i \Phi-(\theta_{\omega_1})_i =0~,
\cr
&&
\partial_a\Phi=0~,~~~
\la{dn=8su2d}
\eea
respectively, where $M,N=0, \dots, 9$ are spacetime coordinate indices and the indices
of the rest of the formulae have been converted to frame indices
using the frame, and $\epsilon_{051627}=1$. The 1-form $\theta_{\omega_1}$ is the Lee
form associated with $\omega_1$
and its definition can be found in \cite{het1}.

The first condition in (\ref{dn=8su2g})  implies that the six vector fields $e_a$ dual
to the 1-forms $e^a$ are Killing and that the twist $de^a$
of these vector fields
is determined in terms of  $H$. This together with $\hat\nabla\omega_r=0$ imply that
${\rm hol}(\hat{\nabla})\subseteq SU(2)$ as expected.
The (\ref{dn=8su2d})
conditions arise from the dilatino Killing spinor equation. In particular,
the first two conditions imply that the Lie brackets of any two of $e_a$ vector fields associated to
the 1-forms $e^a$ close in the span
of the vector fields, ie
the six parallel vector fields span a metric Lorentzian Lie algebra. The structure constants of this
 Lie algebra are the $H_{abc}$ components of $H$.
Moreover, the structure constants are  self dual. The 6-dimensional self-dual Lorentzian Lie algebras have been classified
\cite{josela} and they are isomorphic to
\bea
\bR^{5,1}~,~~~\mathfrak{sl}(2, \bR)\oplus\mathfrak{su}(2)~,~~~\mathfrak{cw}_6~.
\la{liealg}
\eea
In particular, the non-vanishing Lie algebra relations of $\mathfrak{cw}_6$ are
\bea
[{\bf e}_p, {\bf e}_q]=\beta_{pq}\, {\bf k}~,~~~[{\bf t}, {\bf e}_p]=\beta_{pq} \, {\bf e}_q~,~~~\beta_{pq}={1\over2}
\e_{pq}{}^{p'q'} \beta_{p'q'}~,
\la{algcw6}
\eea
where $\beta$ is a constant 2-form  $p,q, p', q'=1,6, 2, 7$, and the Lorentzian metric is
\bea
<{\bf e}_p, {\bf e}_q>=\delta_{pq}~,~~~ <{\bf k}, {\bf t}>=1~.
\eea
The  condition on $\beta$ arises from the requirement
of self-duality of the structure constants of $\mathfrak{cw}_6$.

The conditions on $de^a$ in (\ref{dn=8su2d})  restrict the twist of the parallel vector fields $e_a$
along the $e^i$ directions to be $(1,1)$ with respect
to all three endomorphisms $I_r$. Observe that $de^a_{ij} =H^a{}_{ij}$. These restrictions
will be interpreted  as a anti-self-duality condition for a curvature.
The condition on the Lee form  $\theta$ will also be explained later.

The last condition in (\ref{dn=8su2d}) implies that the dilaton
is invariant under the parallel vector fields. Since the metric is also invariant,
it remains to discuss the invariance of $H$.
The Killing spinor equations alone do not imply that $H$ is invariant. However if
$H$ is closed, then it is straightforward to show
that $H$ is also invariant, ${\cal L}_a H=0$.  Moreover in a perturbation scenario,
assuming that at zeroth order $H$ is invariant, because
it is closed, the corrections to $dH$ are constructed from curvature tensors and
so are invariant. So it is reasonable to choose $H$ to be invariant
as well. It is also the case that
\bea
i_{e_a}\omega_r=0~,~~~{\cal L}_a\omega_r=0~.
\la{invcono}
\eea
The latter condition follows because $\omega_r$ is $\hat\nabla$-parallel and that
$de^a_{ij}$ is (1,1) with respect to all three $I_r$.

\subsection{Solution of the Killing spinor and field equations}

\subsubsection{Solution of the Killing spinor equations}

Taking  $H$ to be invariant under all isometries $e_a$, the conditions in (\ref{dn=8su2g}) and (\ref{dn=8su2d}) have been interpreted
in terms of principal bundles and their connections \cite{het1}. In particular provided that the infinitesimal action generated by
the $\hat\nabla$-parallel vector fields can be integrated to a group action,
the spacetime, $M$, is a {\it principal bundle} $M=P(B, G;\pi)$, where $\mathfrak{ Lie}\, G$ is one of the self-dual algebras
in (\ref{liealg}). It is clear from  (\ref{invcono})
that $\omega_r$ are the pull-backs of 2-forms on $B$ and so one can write  $\omega_r=\pi^*\tilde \omega_r$. As a consequence,
 the base space $B$ is an HKT 4-dimensional
manifold, $(d\tilde s^2_{(4)}, \tilde H)$ equipped with the Hermitian forms $\tilde\omega_r$.   The condition
$\theta=2d\Phi$ implies that
$B$ is {\it conformally balanced}.

In addition, the principal bundle is equipped with a connection $\lambda^a\equiv e^a$, see eg \cite{nomizu}. The curvature
\bea
{\cal F}^a\equiv d\lambda^a-{1\over2} H^a{}_{bc}\, \lambda^b \lambda^c={1\over2} H^a{}_{ij}\, e^i\wedge e^j
\eea
of $\lambda^a$  is (1,1) with respect all the complex structures of the HKT manifold. Now if the orientation\footnote{
The opposite orientation was chosen in \cite{het1, het2}.}
of $B$ is chosen with respect to the $\omega_1$ complex structure,
 then the (1,1) condition on ${\cal F}$ with respect to $I_r$
is compatible with the restriction that ${\cal F}$ is  {\it anti-self-dual},
\bea
{\cal F}^a_{ij}=-{1\over2} \epsilon_{ij}{}^{kl}  {\cal F}^a_{kl}~,~~~
d{\rm vol}={1\over2}\omega_1\wedge \omega_1={1\over 4!}\epsilon_{i_1\dots i_4}
e^{i_1}\wedge \dots\wedge e^{i_4}~.
\eea
Given these data,
the metric and $H$ of
spacetime can be written as
\bea
ds^2&=&\mt_{ab} \lambda^a \lambda^b+\pi^*d\tilde s^2_{(4)}
\cr
H&=&{1\over3} \mt_{ab} \lambda^a\wedge d\lambda^b+{2\over3} \mt_{ab} \lambda^a\wedge {\cal F}^b+\pi^* \tilde H~,
\la{metreH}
\eea
where $\mt$ is a bi-invariant metric on $\mathfrak{ Lie}\, G$. We can always choose the frame $\lambda$
such that  $\mt$ is the standard Lorentzian metric. Observe that $H$ can be written as a sum of the Chern-Simons
form of $\lambda$ and the 3-form on $B$.

The solution to the Killing spinor equations (\ref{metreH}) can be simplified further.
In particular since $B$ is a {\it conformally balanced HKT manifold}, there is a function $h$ of $B$ such that\footnote{Our form conventions are
$\l={1\over k!}\l_{i_1\dots i_k} dx^{i_1}\wedge\dots\wedge dx^{i_k}$ and Hodge dual is taken as
$\star\l_{i_{k+1}\dots i_n}={1\over k!}\l_{i_1\dots i_k} \e^{i_1\dots i_k}{}_{i_{k+1}\dots i_n}$.}
\bea
d\tilde s_{(4)}^2=h\, ds_{\rm hk}^2~,~~~~\tilde H=-\star_{\rm hk} dh~,~~~e^{2\Phi}=h~,
\eea
where $ds_{\rm hk}^2$ is a 4-dimensional hyper-K\"ahler metric.
Since the anti-self-duality condition of the curvature  of $\lambda$ is conformally invariant, $\lambda$ can be thought of as
a anti-self-dual principal bundle connection over the hyper-K\"ahler manifold $B_{hk}$. Thus all such solutions
of the Killing spinor equations can be constructed from a {\it hyper-K\"ahler 4-dimensional manifold $B_{\rm hk}$} and
{\it a anti-self-dual principal bundle connection on $B_{\rm hk}$} with gauge group a 6-dimensional
Lorentzian group $G$ which has a self-dual Lie algebra
 (\ref{liealg}). To summarize, the solution to the Killing spinor equations can be written as in (\ref{xxy}).
To find the explicit solutions, one has to determine the function $h$.

\subsubsection{Solution of the field equations}

It remains to solve the field equations to determine $h$. For this suffices
 to solve the (anomalous)  Bianchi identity for $H$
which can now be written as
\bea
dH=d\pi^*\tilde H+\mt_{ab} {\cal F}^a\wedge {\cal F}^b=-{\a'\over4} {\rm tr} (\check R^2-F^2)+{\cal O}(\a'^2)~.
\la{closeH}
\eea
Taking $H$  to be closed in the zeroth order in $\a'$, the Bianchi identities imply that $\hat R_{AB,CD}=\check R_{CD,AB}$. Since
${\rm hol}(\hat\nabla)\subseteq SU(2)$, the non-vanishing components of $\check R$ are $\check R_{ij, AB}$ and moreover $\check R_{ij, AB}$
is (1,1) with respect to $I_r$.
Similarly, the gaugino Killing spinor equation implies that the non-vanishing components of $F$ are $F_{ij}$ and moreover
$F$ is (1,1) with respect to $I_r$. Using these, we find that
\bea
-\nabla^2_{\rm hk} h-{1\over2} \mt_{ab}\,{\cal F}^a_{ij}\,\, {\cal F}^{bij}={\alpha'\over8}
\big({\rm tr}\check R_{ij}{\check R}^{ij}-{\rm tr}
F_{ij} F^{ij}\big)+{\cal O}(\alpha'^2)~,
\la{fineqn}
\eea
where all indicated contractions of $i,j$ indices are with respect to the hyper-K\"ahler metric.
To find a solution, this equation must be solved for the function $h$.

In what follows,
we shall mostly explore solutions for which the $\hat\nabla$ connection is identified with the
gauge connection $A$. As a result the curvature square terms in the right-hand-side of
(\ref{fineqn}) cancel and there is no anomalous contribution in the Bianchi
identity for $H$, ie $dH=0$. This identification can always be done if the associated backgrounds are thought as solutions
of the common sector of type II supergravities. However some care is required when the
$\hat\nabla$ connection is identified with $A$ in the context of heterotic strings, see also \cite{tseytlin1}
for a similar discussion. This is because
for generic backgrounds ${\rm hol}(\check \nabla)\subseteq SO(9,1)$ and so it cannot be embedded in the
$SO(32)$ or $SO(16)\times SO(16)$ internal symmetry of the fermionic sector of the worldvolume action
necessary for the construction of $Spin(32)/\bZ_2$ and $E_8\times E_8$ heterotic strings, respectively.
In the explicit examples that we shall consider below for the ${\rm hol}(\hat\nabla)\subseteq SU(2)$ class
of backgrounds, ${\rm hol}(\check \nabla)$ is always an orthogonal
compact group which can be embedded in both $SO(32)$ and $SO(16)\times SO(16)$ and so this difficulty
does not arise. We shall give the holonomy of ${\rm hol}(\check\nabla)$ in most examples for completeness.

\newsection{Trivial principal bundle solutions}

\subsection{$B_{\rm hk}=\bR^4$}

\subsubsection{WZW models and the self-dual string}

\vskip 0.2cm
\leftline{{\underline {Solution}}}
\vskip 0.2cm

Suppose that the principal bundle connection is trivial. This implies that ${\cal F}=0$ and so one recovers the
Maurer-Cartan equations
\bea
d\lambda^a={1\over2} H^a{}_{bc}\lambda^b\wedge \lambda^c~.
\eea
So $\lambda$ are the left-invariant vector fields on the fibre group $G$.  Moreover suppose that
$\check R=F=0$, ${\rm hol}(\check\nabla)=\{1\}$,
 so that the contribution from the curvature square terms in (\ref{fineqn}) vanishes. The resulting equation gives
\bea
\nabla^2_{\rm hk} h=0~.
\la{hypharm}
\eea
One solution of this is to take $h$ constant, eg $h=1$. In such a case, one finds the solution
\bea
ds^2=\mt_{ab} \lambda^a \lambda^b+ds^2(\bR^4)~,~~~H={1\over6}\, H_{abc} \lambda^a\wedge \lambda^b\wedge \lambda^c~,~~~\Phi={\rm const.}
\eea
For the three choices of self-dual Lie algebras (\ref{liealg}), one finds the Minkowski vacuum solution, $\bR^{9,1}$,
$AdS_3\times S^3\times \bR^4$ and $CW_6\times \bR^4$ up to discrete identifications, respectively.
Note that $SL(2,\bR)=AdS_3$ and $SU(2)=S^3$.
All these solutions have constant dilaton.

The $AdS_3\times S^3\times \bR^4$ solution can be identified as the lifting
of the near horizon geometry of the self-dual string to heterotic supergravity. The radius of $AdS_3$ is equal to that of $S^3$
because of the self-duality condition. We shall use this later in the M-theoretic interpretation of 5-brane solutions.

The invariant metric on any $CW$ space can be written as
\bea
ds^2(CW)=2 dv \big(du+{1\over 8} {\bf y}^t \b^2 {\bf y}\,dv \big)+ d{\bf y}^t d{\bf y}~.
\eea
The solution requires that   $\b$ is self-dual 2-form. So it can be written as a sum of the three self-dual forms on $\bR^4$. As a result,
\bea
ds^2(CW_6)=2 dv \big(du-{\mu^2\over 8} {\bf y}^t  {\bf y}\, dv \big)+ d{\bf y}^t d{\bf y}~.
\eea
where $\mu$ is the length of $\b$.  In fact the $CW_6\times \bR^4$ solution preserves 12 supersymmetries \cite{josehet} and it is
 the Penrose limit of $AdS_3\times S^3\times \bR^4$ \cite{plads}.

\vskip 0.2cm
\leftline{{\underline {Flux quantization}}}
\vskip 0.2cm

We shall examine the worldvolume theory of strings in such backgrounds in more detail later.
Here we remark that since the relative radius  of $AdS_3$ and $S^3$  in the $AdS_3\times S^3\times \bR^4$
background is fixed because
of the self-duality condition, there is only one overall coupling constant. Moreover appealing
to a Dirac quantisation condition
\bea
\int_{S^3} H=k~,
\la{kquant}
\eea
is an integer, $k\in\bZ$, in some units.

\subsubsection{5-branes with flat and curved worldvolume}

\vskip 0.2cm
\leftline{{\underline {Flat worldvolume}}}
\vskip 0.2cm
There are three types of 5-brane solutions with different worldvolume geometry for each choice
of Lie algebra (\ref{liealg}).
The solution found in \cite{callan} corresponds to the case that  $\mathfrak{Lie}\,G=\bR^{5,1}$.
Allowing for delta function sources and identifying the $\check\nabla$ connection with the gauge connection $A$,
 one finds that a solution\footnote{Observe that $h+b_i x^i$ is also a harmonic function if $h$ is. So there
 is a more general solution which however is not asymptotically Minkowski.}   of (\ref{hypharm}) is
\bea
h=1+\sum_\ell{Q_\ell\over |x-x_\ell|^2}~.
\la{harm}
\eea
Observe that ${\rm hol}(\check \nabla)\subseteq SO(4)$.
This solution can be rewritten as
\bea
ds^2=ds^2(\bR^{5,1})+h\, ds^2(\bR^4)~,~~~H=-\star dh~,~~~e^{2\Phi}=h
\la{s5brane}
\eea
and it is the multiple 5-brane solution of the heterotic string. The 5-branes are located at the $x_\ell$ positions
in $\bR^4$, and $\bR^{5,1}$ is the worldvolume.

The solution has two asymptotic regions. As $|x|\rightarrow \infty$, the spacetime approximates the Minkowski vacuum. On the other
has as $|x-x_\ell|\rightarrow 0$, ie close to the position of a 5-brane, the solution becomes the WZW model $\bR^{5,1}\times S^3\times \bR$
with linear dilaton. The dilaton is infinite at $x=x_\ell$, ie string theory is strongly coupled, but this point lies infinite affine distance
way
from any other point of the spacetime.

The 5-brane charge $p$, per unit worldvolume, is
\bea
p={1\over {2\rm Vol} (S^3)}\int_{S_\infty^3} H
\la{coninf}
\eea
where ${\rm Vol} (S^3)=2\pi^2$ is the volume of unit round $S^3$ and the integral is over the $S_\infty^3\subset \bR^4$ at infinity, $|x|\rightarrow \infty$,
of the transverse $\bR^4$
to the worldvolume directions
of  the 5-brane.

A straightforward computation reveals that for the solution (\ref{s5brane}), the 5-brane charge is
\bea
p=\sum_\ell Q_\ell~.
\eea
 Appealing to the Dirac-like quantization condition, $p$ must also be quantized in some units.

\vskip 0.4cm
\leftline{{\underline {Curved worldvolume}}}
\vskip 0.2cm

There are  two generalizations of the multiple 5-brane solution by replacing the worldvolume $\bR^{5,1}$ geometry
of the above solution
with either $G=AdS_3\times S^3$
or $G=CW_6$. These solutions are
\bea
ds^2=ds^2(G)+h\, ds^2(\bR^4)~,~~~H={1\over6}\, H_{abc} \lambda^a\wedge \lambda^b\wedge \lambda^c-\star dh~,~~~e^{2\Phi}=h~,
\la{wrap5b}
\eea
where $\check\nabla$ is again identified with the gauge connection $A$, ${\rm hol}(\check\nabla)\subseteq SO(4)$, and
$h$ is  given in (\ref{harm}). These solutions have again two asymptotic regions.
 As $|x|\rightarrow \infty$, these solutions become either $AdS_3\times S^3\times \bR^4$
or $CW_6\times \bR^4$
with constant dilaton.  While as $|x-x_\ell|\rightarrow 0$, ie near the position of a 5-brane,
 these solutions become either $AdS_3\times S^3\times S^3\times \bR$ or $CW_6\times S^3\times \bR$
with linear dilaton. The string coupling becomes infinite at the position of the 5-brane but this position
is at infinite affine distance away.
String theory is again strongly coupled at the position of the 5-brane.

The solution (\ref{wrap5b})  has the properties of a 5-brane with curved $AdS_3\times S^3$ or $CW_6$ worldvolume.
This is supported from the asymptotic conditions. The 5-brane charge per unit worldvolume is again $p=\sum_\ell Q_\ell$
for the (\ref{wrap5b}) solution and $h$ as in (\ref{harm}). Observe that the size of the worldvolume geometry does not depend on
the transverse coordinate $x$. In the $AdS_3\times S^3$ case, one may also consider the flux (\ref{kquant})
which is again quantized in some units.

\subsection{General hyper-K\"ahler $B_{\rm hk}$}

\subsubsection{WZW models and compactifications}

It is clear from (\ref{hypharm}) that $h$ can taken to be a constant for any hyper-K\"ahler 4-manifold $B_{\rm hk}$.
As a result, one can find the
solutions
\bea
\bR^{5,1}\times B_{\rm hk}~,~~~ AdS_3\times S^3\times B_{\rm hk}~,~~~ CW_6\times B_{\rm hk}~.
\eea
Choosing $B_{\rm hk}=K_3$ or $T^4$, $\bR^{5,1}\times K_3$  and $\bR^{5,1}\times T^4$ can be
interpreted as the vacua of a $K_3$ and $T^4$ compactification of the
heterotic string to six dimensions, respectively. Similarly $AdS_3\times S^3\times K_3$ and  $AdS_3\times S^3\times T^3$
can be interpreted as the vacua of an $AdS_3$   compactification
of the heterotic string on $S^3\times K_3$ and $S^3\times T^4$, respectively.
In all the above
backgrounds the $\check\nabla$ connection is identified with the gauge one $A$
and ${\rm hol}(\check\nabla)\subseteq SO(4)$.
Of course $B_{\rm hk}$ can be chosen to be non-compact. In such a case
for any choice of a 4-dimensional hyper-K\"ahler manifold, we get a new solution.

\subsubsection{5-branes with flat and curved worldvolume  in hyper-K\"ahler manifolds}

The solution (\ref{wrap5b}) can be  generalized by replacing $\bR^4$ with any other 4-dimensional
hyper-K\"ahler manifold $B_{\rm hk}$. In such a case, $h$ satisfies (\ref{hypharm}) where the harmonic condition is with respect
to the  hyper-K\"ahler metric on $B_{\rm hk}$. The resulting solution is
\bea
ds^2=ds^2(G)+h\, ds^2(B_{\rm hk})~,~~~H={1\over6}\, H_{abc} \lambda^a\wedge \lambda^b\wedge \lambda^c-\star_{\rm hk} dh~,~~~e^{2\Phi}=h~,
\la{hypwrap5b}
\eea
where $G=\bR^{5,1}$, $AdS_3\times S^3$ or $CW_6$.
This solution has the interpretation as 5-branes located on a 4-dimensional
hyper-K\"ahler manifold with worldvolume geometry $G$. However, the explicit expression for $h$ depends
on the choice of hyper-K\"ahler metric, and so
explicit solutions can be given only on a case by case basis. In all these cases the $\check\nabla$ connection
can be identified with the gauge one in the context of heterotic string since ${\rm hol}(\check\nabla)\subseteq SO(4)$.
 It is straightforward to construct explicit solutions in some cases. For example an explicit solution can be found
if  the hyper-K\"ahler metric is chosen to be
the Gibbons-Hawking metric.

\newsection{Non-trivial principal bundle solutions}

\subsection{Principal bundle connections}

Before we proceed to construct solutions with an abelian and a non-abelian principal bundle connection $\lambda$, we shall first express
 $\lambda$ in terms of  standard vector bundle connection $C$. This strictly is not necessary since principal bundle connection theory
  is well established \cite{nomizu}. However, the analysis may look more familiar in terms of vector bundle connections.
  Indeed  given $C$ a vector bundle connection,  write
the local expression
\bea
\lambda=-g^{-1}d g+ g^{-1} C g
\la{princon}
\eea
where $g$ is a function of the principal bundle with values on the gauge group $G$.  Since  $C$
takes values in the Lie algebra of the gauge group, it is clear that $\lambda$ is also $\mathfrak{Lie}\, G$-algebra valued.
Moreover as required, it transforms under the adjoint representation of $\mathfrak{Lie}\, G$
under the right action of $G$ on the spacetime. Restricting  $\l$  on the fibre $G$, one recovers the
left-invariant 1-forms of the group $G$.
Next, the curvature
\bea
{\cal F}=d\l-\l\wedge \l
\eea
of $\lambda$ is
\bea
{\cal F}=g^{-1} (dC-C\wedge C) g=g^{-1} F(C) g~,
\eea
where $F(C)$ is the curvature of the connection $C$. Therefore  the curvature of the principal bundle ${\cal F}$ is related to the
 curvature of connection $C$ up to a gauge transformation.
Observe that if $G$ is abelian, then (\ref{princon}) can be simplified. For example if $G=U(1)$, taking $g=e^{-i\theta}$, then
$\l=i(d\theta+C)$, where $C$ is locally a real 1-form. Such principal bundle connections appear in black hole solutions with rotation
or brane solutions with wrapping on a torus.

 It is now clear that since  the metric $\mt$  on $\mathfrak{Lie}\, G$ is bi-invariant, one can replace ${\cal F}$ in the the equations (\ref{closeH}) and
 (\ref{fineqn}) with $F(C)$. The gauge transformation $g$ is eliminated from gauge invariant expressions like the first Pontryagin class.
 So to find explicit solutions, it is sufficient to begin from a anti-self-dual   connection $C$
 and then use the formula (\ref{princon}) to construct the principal bundle connection.

As a final remark on the relation between principal bundle $\lambda$ and vector bundle  $C$ connections,
it is customary in some parts of the physics literature, instead of using $\lambda$,
to use
\bea
\rho\equiv g \lambda g^{-1}=-dg g^{-1}+C
\eea
where $dg g^{-1}$ are identified with the {\it right-invariant} 1-forms of $G$. Moreover typically  $dg g^{-1}$ are explicitly
expressed in terms of the coordinates of $G$. Since $\rho$ and $\lambda$ are related up to a conjugation
all quantities of $\lambda$ can be expressed equally well in terms of $\rho$. However, we shall
continue the analysis in terms of $\lambda$ because the geometric interpretation of both the metric of the spacetime and the 3-form
field strength is more transparent in terms of $\lambda$.

\subsection{Non-abelian connections and $B_{\rm hk}=\bR^4$}

\subsubsection{Instantons}

 Before we proceed to explicitly construct supergravity solutions   for $B_{\rm hk}=\bR^4$, we shall summarize some properties
 of $SU(2)$ instantons. In fact the relevant instanton solutions are those for the gauge group   $SL(2,\bR)\times SU(2)$. It is not apparent though that
 there are non-abelian anti-self-dual connections for the gauge group $SL(2,\bR)$, so we shall consider non-abelian connections only for the $SU(2)$
 subgroup.

 Such connections can be constructed using the t'Hooft ansatz which can be described as follows.
 On $\bR^4$ there are two commuting (constant) hyper-complex structures $I_r$ and $J_r$, ie
 \bea
 I_r I_s=-\delta_{rs} {\bf 1}_{4\times 4}+\e_{rst} I_t~,~~~J_r J_s=-\delta_{rs} {\bf 1}_{4\times 4}+\e_{rst} J_t~,~~~[I_r, J_s]=0
 \eea
 which can be constructed from a basis of self dual and anti-self-dual 2-forms, respectively. It is well known
 that $\mathfrak{so}(4)=\mathfrak{so}(3)\oplus \mathfrak{so}(3)$ and $I_r$ and $J_r$ span the two $\mathfrak{so}(3)$
 subalgebras, respectively. Next consider $\bR^4$ as a hyper-K\"ahler manifold with respect to $I_r$, and write
 \bea
 C^r_i=(I_r)^j{}_i \partial_j \log f~,
 \la{inst}
 \eea
 for some function $f$ of $\bR^4$ which must be determined.
The anti-self-duality condition on $F(C)$, which is equivalent to requiring that $F(C)$ is (1,1) and traceless with respect to $I_1$, implies that
\bea
{1\over f} \,\partial^2 f=0
\eea
Thus a solution for $f$ is
\bea
f=1+\sum^\nu_{\ell=1} {\rho^2_\ell\over |x-x_\ell|^2}~,~~~x\in \bR^4~,
\la{sinst}
\eea
This is the $\nu$-instantons solution with the $\ell$-th  instanton located at the pole  $x_\ell$ and   $\rho^2_\ell$
 is identified  with its size.

These solutions though do not have the correct asymptotic behaviour and are singular. In particular $F(C)$ diverges at the positions of the instantons.
However after a singular gauge transformation, the 1-instanton  configuration can be rewritten as
\bea
C^r_i= 2(J_r)_{ij} {x^j\over |x|^2+\rho^2}~,
\la{onein}
\eea
where translation invariance has been used to locate the instanton at the origin in $\bR^4$.

The instanton configurations are characterized by their first Pontryagin number or instanton number,
\bea
\nu=-{1\over 16\pi^2}\int \delta_{rs}\,F^r\wedge F^s~.
\eea
In particular, for the 1-instanton configuration $\nu=1$.

\subsubsection{Wrapped 5-branes on $S^3$}

To find the solutions associated with the non-trivial $SU(2)$ principal bundle, we begin by considering
the instanton solution (\ref{inst}) for $f$ given in (\ref{sinst}). A direct computation reveals that
\bea
-{1\over2} \delta_{rs}\,{\cal F}^r_{ij}\,\, {\cal F}^{sij}=-{1\over 2} \delta_{rs} F^r(C)_{ij} F^{s}(C)^{ij}=\partial^2 \partial^2 \log f~.
\eea
Substituting this into the equation for $h$ (\ref{fineqn}), observing that ${\rm hol}(\check \nabla)\subseteq SO(7)$,
 and identifying the $\check\nabla$ with the connection $A$, we have that
\bea
-\partial^2 h+\partial^2 \partial^2 \log f=0~.
\eea
Clearly, if one allows for delta function sources, then $h$ is not uniquely determined in terms of $f$. In particular,
one finds that
\bea
h=h_0+\partial^2\log f~,~~~h_0=s+\sum_n {Q_n\over |x-x_n|^2}~,
\la{gsol}
\eea
where $s=1$ or $s=0$.

Let us investigate the one instanton case with size $\rho^2$ in more detail.  To begin consider the special case for which
\bea
h=1+{4\over |x|^2}+\partial^2\log f=1+4 {|x|^2+2\rho^2\over (|x|^2+\rho^2)^2}~,~~~f=1+{\rho^2\over |x|^2}~,
\la{xxx}
\eea
receives contributions both from the  instanton, $f$,  and $h_0$.
This is precisely the expression for $h$ that one would find if the smooth instanton solution (\ref{onein})
was used to calculate the Pontryagin class.

The resulting supergravity solution given by substituting (\ref{xxx}) in (\ref{xxy}) is  smooth.
Moreover it   has one asymptotic region $AdS_3\times S^3\times \bR^4$ as $|x|\rightarrow \infty$.
Integrating $H$ on the asymptotic 3-sphere  $S_\infty^3\subset \bR^4$,
one detects 5-brane charge
\bea
 p=4~.
 \eea
So the smooth $\nu=1$ instanton gives rise to 4 units of 5-brane change. It is not apparent why this is the case
but this result arises from the standard normalization for 5-brane and instanton charges.

Furthermore the  geometry of the solution associated with (\ref{xxx})
at  $|x|\rightarrow 0$ is again $AdS_3\times S^3\times \bR^4$ but this region is not asymptotic. This is because
it can be reached in finite proper distance from any other region of spacetime apart
from that for which $|x|\rightarrow \infty$.
The dilaton is bounded on the spacetime and so there are no strongly coupled regions. These asymptotic conditions are reminiscent
of those of dyonic 5-brane backgrounds \cite{zamaklar} which preserve 1/4 of supersymmetry.

It is clear that the spacetime of the  solution (\ref{xxy}) for $h$ given in (\ref{xxx}) is a product $M=AdS_3\times X_7$, where $X_7$ is a
non-trivial  $SU(2)$-principal bundle over
$\bR^4$. The size of the $S^3$ fibre is the same at every point of $\bR^4$. The isometries of $X_7$ are
$SU(2)\times SO(4)$.  So the symmetry group of the whole solution is
 $SL(2,\bR)\times SL(2,\bR)\times SU(2)\times SO(4)$.
Observe that if the instanton connection vanishes and $X_7=S^3\times \bR^4$, the isometry
group is $SU(2)\times SU(2)\times SO(4)$. The presence of an instaton breaks some of the isometries
of the background.

Adding further 5-branes  to (\ref{xxx}) as
\bea
h=1+4 {|x|^2+2\rho^2\over (|x|^2+\rho^2)^2}+ {Q\over |x|^2}~,
\la{xxx5}
\eea
one finds another
solution with the same asymptotic geometry at $|x|\rightarrow \infty$. But now at the position  of the 5-brane
there is another
asymptotic region with geometry $AdS_3\times S^3\times S^3\times \bR$ and a linear dilaton. The brane charge detected at
$|x|\rightarrow \infty$ is $p=4+Q$. This solutions has the same symmetries as that given for $h$ in (\ref{xxx}).
Of course (\ref{xxx5}) can be further generalized by adding additional 5-branes located at different positions
in $\bR^4$.

\subsubsection{ADHM construction and 5-branes}

One can generalize the previous example to multi-instanton solutions using the ADHM construction \cite{adhm}.
The simplest case to describe is that of $\nu$ instantons for the gauge group $Sp(r)$. For this
consider a $(\nu+r)\times r$ matrix $V$ of quaternions  which satisfy the linear relation
\bea
\Delta^\dagger V=0
\eea
where
\bea
\Delta=a+bx~,
\eea
$a$ and $b$ are $(\nu+r)\times \nu$ matrices of quaternions, $x\in \bH=\bR^4$ are the coordinates of the
base space $\bR^4$ and the conjugation indicated is matrix transpose followed by quaternionic conjugation.
The gauge potential is $C=V^\dagger dV$, where the differentiation is with respect to $x$.
After some computation, one finds that
\bea
-{1\over2} \delta_{rs}\,{\cal F}^r_{ij}\,\, {\cal F}^{sij}=-{1\over 2} \delta_{rs} F^r(C)_{ij} F^{s}(C)^{ij}
=\partial^2 \partial^2 {\rm Tr}\log {\rm f}~,
\la{rmf}
\eea
where ${\rm f}$ is the $\nu\times \nu$ matrix
\bea
{\rm f}=\Delta^\dagger \Delta=a^\dagger a+a^+bx+x^\dagger b^\dagger a+x^\dagger b^\dagger bx~.
\la{del}
\eea

To construct new solutions (\ref{xxy}) based on the $SU(2)$ $\nu$-instanton connection given by the
ADHM construction, substitute  (\ref{rmf}) in (\ref{fineqn})to find
\bea
h=s+\partial^2 {\rm Tr}\log  {\rm f}~,~~~s=0,1~,
\la{multisol}
\eea
This is the multi-instaton generalization of (\ref{xxx}), $s=1$.
The 1-instanton case for the gauge group $SU(2)=Sp(1)$ that we have given in (\ref{xxx}) can be recovered
for
\bea
\Delta^\dagger=(\rho,  x^\dagger)
\eea
where $\rho$ is a quaternion with a real only component. Observe that the ADHM construction gives
the smooth instanton solutions with the expected asymptotic conditions without the need of a singular
gauge transformation.

It is likely that the supergravity solution (\ref{xxy}) associated with  (\ref{multisol})
 is smooth and the dilaton is bounded
everywhere on spacetime. However, we have not been able to show that $h>0$ for $\nu>1$.
As $|x|\rightarrow \infty$ and for $s=1$, the spacetime becomes $AdS_3\times S^3\times \bR^4$.
The 5-brane charge detected at infinity, $s=1$, is $p=4\nu$.
 If instead $s=0$, as $|x|\rightarrow \infty$, the spacetime becomes $AdS_3\times S^3\times S^3\times \bR$. This
 is a linear dilaton background.

The solution (\ref{xxy}) associated with in (\ref{multisol})
has continuous
moduli which can be identified with that of $SU(2)$ instantons. In particular,
it  will depend
on $8\nu-3$ parameters.
The spacetime is again a product $SL(2,\bR)\times X_7$. For a generic $\nu$ instanton connection, the isometry
group of $X_7$ is $SU(2)$. So the symmetry group of the whole background is $SL(2,\bR)\times SL(2,\bR)\times SU(2)$.
Enhancement of symmetry can occur if the positions of the instantons and their relative orientations are chosen
appropriately.

A more general solution can be constructed by adding 5-branes located at different points in $\bR^4$. In particular,
one can solve (\ref{fineqn}) as
\bea
h=s+\partial^2 {\rm Tr}\log  {\rm f}+\sum^\kappa_{n=1} {Q_n\over |x-x_n|^2}~,~~~s=0,1~,
\la{multisol5}
\eea
where ${\rm f}$ is given in (\ref{del}).
Apart from the asymptotic regions we have mentioned above as $|x|\rightarrow \infty$ for (\ref{multisol}), this new solution
has new asymptotic regions at the positions of the 5-branes $|x-x_n|\rightarrow 0$. The geometry
of the spacetime near the positions of the 5-branes becomes $AdS_3\times S^3\times S^3\times \bR$. This is
a linear dilaton background. The 5-brane charge detected at infinity, $|x|\rightarrow \infty$, $s=1$, is
$p=4\nu+\sum_n Q_n$.

 The solution (\ref{xxy}) associated with in (\ref{multisol5})
has continuous
moduli that of $\nu$  $SU(2)$ instantons. In addition the positions of the 5-branes are new moduli. So
it depends of $8\nu+4\kappa-3$ parameters. The symmetry group of a generic solution
is again $SL(2,\bR)\times SL(2,\bR)\times SU(2)$.

\subsubsection{5-brane probes}

As further evidence for the interpretation of (\ref{xxx}), we can do a probe computation. In particular consider a 5-brane probe
which is embedded in spacetime by identifying  the worldvolume probe coordinates  with the
worldvolume coordinates $AdS_3\times S^3$ of the solution
(\ref{xxx}). Moreover, take the transverse coordinates $x$ of the spacetime
to be constant. The induced metric
on the probe worldvolume is that the $AdS_3\times S^3$. Moreover such a probe is supersymmetric. This follows immediately
from the kappa-symmetry condition \cite{duff, becker, kallosh}
\bea
\Gamma\e=\e~,~~~
\eea
where $\Gamma$
 is the kappa symmetry projector of the 5-brane and $\e$ is a Killing spinor.
For the embedding described above, the kappa symmetry condition gives
\bea
\Gamma_{051627}\e=\e~.
\eea
All Killing spinors of the supergravity background (\ref{xxx}) satisfy
this kappa-symmetry condition and so the embedding is supersymmetric.

\subsubsection{Abelian connections}

The construction described above for non-abelian connections can be easily extended to abelian ones. If $B_{\rm hk}=\bR^4$, abelian
anti-self-dual connections can be constructed by writing
\bea
C_i=(I_r)^j{}_i \partial_j  f^r
\eea
and requiring that $f^r$ are harmonic functions. Such solutions typically will be singular at the positions of the
harmonic functions $f^r$. One can make superposition of abelian and non-abelian gaugings. For example,
one can consider principal bundles with a non-abelian $SU(2)$ connection but only gauging an abelian subgroup
of $SL(2,\bR)$. The gauging of a time-like direction in $SL(2,\bR)$ is of particular interest because
it allows the construction of smooth solutions for the cases that $B_{\rm hk}$ is a compact hyper-K\"ahler manifold.
However, this may lead to a Lorentzian holonomy for $\check \nabla$.

\subsubsection{General hyper-K\"ahler manifolds}

It is clear that the construction of solutions presented above can be generalized to any 4-dimensional
hyper-K\"ahler manifold $B_{\rm hk}$. Let us consider the case that $B_{\rm hk}$ is compact. To find
a solution, one has to solve the equation
\bea
\nabla^2_{\rm hk} h=-{1\over2} \mt_{ab}\,{\cal F}^a_{ij}\,\, {\cal F}^{bij}
\la{compcon}
\eea
for $h$. A partial integration formula easily reveals that there is no smooth solution for $h$
unless
\bea
\int_{B_{\rm hk}} \mt_{ab}\,{\cal F}^a\,\wedge\, {\cal F}^{b}=0~.
\eea
Assuming that we have only an non-trivial $SU(2)$ connection, this condition cannot be met because the
integral is proportional to the instanton number which is not zero.

To construct smooth solutions
the contribution of the $SU(2)$ instanton must be canceled. For this, we allow for rotation, ie for an abelian
connection along a time-like direction in $SL(2,\bR)$. Since $\mt$ is a Lorentzian metric, the consistency
condition is now written as
\bea
-\int_{B_{\rm hk}} \,{\cal F}^0\,\wedge\, {\cal F}^{0}+\sum_{r=1}^3\int_{B_{\rm hk}} \,{\cal F}^r\,\wedge\, {\cal F}^{r}=0~.
\la{intcon}
\eea
Because of the relative minus sign, the $U(1)$ instanton can be chosen to cancel the contribution from the
$SU(2)$ instantons. However now the holonomy of $\check\nabla$ may be Lorentzian and so  $\check\nabla$
cannot be identified with the gauge connection in the context of the heterotic string. In such a case,
the contribution of the curvature square terms in (\ref{fineqn}) should be taken into account.
The condition for the existence of a $U(1)$ instanton connection
is
\bea
\int_{B_{\rm hk}} {\cal F}^0\wedge \omega_{\rm hk}=0~.
\eea
This can be generalized to the $SU(2)$ instantons and involves the notion of stable vector bundles \cite{donaldson}.
In addition of course
(\ref{intcon}) must be  satisfied. Alternatively, one can add appropriate sources in the right hand side of
(\ref{compcon})  (anti 5-branes) to cancel the contribution from the $SU(2)$ instantons.

\newsection{Type II and M-theory solutions}

\subsection{Type IIB}
All solutions that we have presented here for which $dH=0$ can be considered as solutions
of type II supergravity theories as well. Generic such solutions in type II will preserve $1/4$ of the supersymmetry,
ie 8 supercharges. However special  solutions can preserve more supersymmetry.

In type IIB, the generic solutions we have constructed  have the interpretation as wrapped NS5-branes on
$G$ with transverse space
any 4-dimensional hyper-K\"ahler manifold. After an S-duality, the generic solution becomes a wrapped D5-brane
on $G$ with transverse space any 4-dimensional hyper-K\"ahler manifold. In particular, the S-dual of the solution
given in (\ref{xxy}) is
\bea
ds^2&=& h^{-{1\over2}} \eta_{ab} \l^a \l^b+h^{1\over2} ds^2(B_{hk})~,~~~~
e^{2\Phi}=h^{-1}~,
\cr
H_{RR}&=&{1\over3} \mt_{ab} \lambda^a\wedge d\lambda^b+{2\over3} \mt_{ab} \lambda^a\wedge {\cal F}^b-\star_{hk} dh~,
\eea
where now $H_{RR}$ is the Ramond-Ramond  3-form field strength.

\subsection{Type IIA and M-theory}

\subsubsection{Intersecting branes}

To give an M-theoretic interpretation of the solutions we have constructed, we shall first explain some
aspects of intersecting branes, see
\cite{paptown} for the supergravity solutions. It is not apparent what the structure is of a supergravity solution which
has the interpretation as a fully localized
brane intersection\footnote{Ansatze for the construction of supergravity solutions with an interpretation as brane intersections
were initially proposed in \cite{paptown}. For a more recent detailed account see \cite{lunin} where
 indirect arguments
and physical intuition, but not a systematic construction,   are used   to find  solutions.}.
A minimal expectation is that one should be able to locate the branes as well as their asymptotic
regions that are involved in the
intersection from data provided by the solution. For simplicity and for the application that we have in mind, suppose
one wants to describe the fully
localized M2-brane intersection with an M5-brane.  In particular, it has been argued in \cite{strom2} that M2-branes
end on M5-branes on a self-dual string\footnote{A partially  localized solution with the interpretation of a M2-brane intersecting a M5-brane
on a self-dual string can be found in \cite{tseytlin}.}. One way to fulfill the above stated  requirements
is to be able to define  two distance functions $\delta_{M2}\geq 0$ and $\delta_{M5}\geq 0$ on the spacetime
which describe the distances
of spacetime points from the location $\delta_{M2}=0$ and $\delta_{M5}=0$ of the M2 and M5 branes, respectively.
 In addition
consider
\bea
\delta_{M2\perp M5}=\sqrt{\delta_{M2}^2+\delta_{M5}^2}~,~~~~\Delta={\delta_{M2}\over \delta_{M5}}~.
\eea
One can then give the following definitions of some regions of  spacetime as follows:
\begin{itemize}

\item Near horizon region of the self-dual string: $\delta_{M2\perp M5}$ small, $\Delta$ finite

\item Overall transverse: $\delta_{M2\perp M5}$ large, $\Delta$ finite

\item Near horizon of the M2-brane: $\delta_{M2}$ small,  $\Delta<<1$

\item Near horizon of the M5-brane: $\delta_{M5}$ small, $\Delta^{-1}<<1$

\end{itemize}
There are other regions that one can define like those that are asymptotic  to either M2- or M5- branes. Under the assumptions
we have made, one finds that apart from the near horizon geometry of the common intersection there are other
geometries that appear in the fully localized M2- with M5-brane intersection, like that of the near M2-brane geometry. This
is simply the region of spacetime which is near the location of the M2-brane. Similarly, one can argue for a near M5-brane
geometry.

\subsubsection{Near $M2$-brane geometry and brane intersections}

The background  (\ref{xxy}) viewed as solution of IIA supergravity can be easily lifted
to 11-dimensional supergravity. In particular, the the 11-dimensional background associated with (\ref{xxy})
for $h$ given in (\ref{xxx}) is
\bea
ds^2_{(11)}&=& h^{-{1\over3}}\,\, \eta_{ab} \l^a\l^b+ h^{{2\over3}} [ds^2(\bR^4)+dy^2]
\cr
F&=& H\wedge dy~,
\la{m2m5}
\eea
where $y$ is the 11th coordinate. Recall that for this solution $G=AdS_3\times S^3$.
The claim is that (\ref{m2m5}) describes the near M2-brane geometry  of an M2 brane
ending on an M5-brane on the self-dual string.  We choose  $\delta_{M5}=r$. As $r<<1$, $\delta_{M5}$ becomes
of the order of $\delta_{M2}$, ie $\Delta$ is finite and $\delta_{M2\perp M5}<<1$. This region is the near horizon
region of the self-dual intersection. The geometry of the spacetime in this region is $AdS_3\times S^3\times \bR^5$
which lies in a finite affine distance from the non-asymptotic points of the spacetime.
This is the expected geometry of the self-dual strings that lie at the intersection.

To continue, we take the 11th coordinate as one of the worldvolume coordinates of the M2 brane which
is transverse to the M5-brane.
Since this
is claimed
to be a near M2-brane geometry, one may expect to find some signs of the $AdS_4\times S^7$
near horizon geometry of the M2-brane. Now the $AdS_3$ part of the spacetime is expected to include
 the distance function $\delta_{M2}$
of the M2-brane as a coordinate. So
the remaining 3 coordinates of $S^3$ and the 4 coordinates of $\bR^4$ must give a description of $S^7$.
This seems to be the
case because  $S^7$ can be viewed as fibration $S^3\rightarrow S^7\rightarrow S^4$.
Moreover this is the instanton bundle, ie the curvature of a principal bundle connection is
anti-self-dual and has instaton number 1. Of course both the $AdS_4$ and $S^7$ geometries of the
near horizon geometry of the M2-brane are distorted because of the presence of the M5-brane.

We do not have a proof that (\ref{m2m5}) is the most general solution of this type. A concern is
that (\ref{m2m5}) is delocalized in the $y$ direction. So there may be a more general solution
of 11-dimensional supergravity which describes the near M2-brane geometry of an M2-brane ending  on
an  M5-brane.
However such solution will have more active fields and it will not be a solution
of the type II common sector alone.

\newsection{Worldvolume theory}

\subsection{Gauged sigma models with WZ term}

To understand the worldvolume theory of the ${\rm hol}(\hat\nabla)\subseteq SU(2)$ class of solutions and in
particular the presence of
the Pontryagin class
of the principal bundle connection in the expression for $dH$, we shall summarize briefly the gauging
of a sigma model with a WZ term \cite{cmhbs}, for the supersymmetric case see \cite{gpbs}.
The gauging which is relevant for the ${\rm hol}(\hat\nabla)\subseteq SU(2)$
class of solutions is that for which the sigma model target space is the fibre $G$ of the principal bundle spacetime
$M=P(G, B;\pi)$.

To begin,  we consider a sigma model  with target space $N$ equipped with metric $g$ and a Wess-Zumino
term $W$, $dW=0$. Moreover, we assume that $N$  admits a group action which leaves  both $g$ and $W$ invariant.
In the standard
analysis for the gauging of such a system,
the worldvolume dimension is two. Here, we shall weaken this assumption, and we
shall take the worldvolume dimension to be $n$.

The action before gauging is
\bea
S={1\over2} \int_\Sigma d^n\sigma g_{pq}\, \partial_\mu X^p\partial^\mu X^q+
\int_{D\subset \Sigma} W_{pqr}\e^{\mu\nu\rho} \partial_\mu X^p
\partial_\nu  X^q \partial_\rho X^r~,
\la{sigmact}
\eea
where $\Sigma$ is the worldvolume and $D$ is a 3-dimensional subset of $\Sigma$.

To begin, let us consider the minimal coupling of $W$ term. This gives
\bea
W={1\over 6} W_{pqr} (dX^p+C^{a_1} \xi_{a_1}^p) \wedge (dX^q+C^{a_2} \xi_{a_2}^q)\wedge (dX^r+C^{a_3} \xi_{a_3}^r)~,
\eea
where $\xi$ are the Killing vector fields that also leave $W$ invariant, and $C$ is a connection which gauges the isometries.
Taking the exterior derivative of $W$, one finds
\bea
dW= {1\over2}\xi_a^p H_{pqr} F(C)^a\wedge (dX^q+C^{a_2} \xi_{a_2}^q)\wedge (dX^r+C^{a_3} \xi_{a_3}^r)~.
\eea
To preserve closure, one has to add non-minimal coupling as
\bea
W&=&{1\over 6} W_{pqr} (dX^p+C^{a_1} \xi_{a_1}^p) \wedge (dX^q+C^{a_2} \xi_{a_2}^q)\wedge (dX^r+C^{a_3} \xi_{a_3}^r)
\cr
&-&
 u_{ap} (dX^p+C^{b} \xi_{b}^p)\wedge F(C)^a
 \la{gfwz}
\eea
where
\bea
i_{\xi_a} W=du_a~.
\eea
Taking again the exterior derivative, one finds that
\bea
dW=- c_{ab} F(C)^a\wedge F(C)^b~,~~~~c_{ab}=\xi^M_{(a} u_{b)M}
\la{anomeqn}
\eea
It can be shown that $c$ is constant. If $c\not=0$, the gauging is considered anomalous. For more details
on the anomalies that arise in the gauging of the Wess-Zumino term of sigma models see \cite{cmhbs} and for their
relation to equivariant cohomology see \cite{josestaciu}.

 It is clear that
(\ref{anomeqn}) is reminiscent to that of (\ref{closeH}).
To make a connection with the ${\rm hol}(\hat\nabla)\subseteq SU(2)$  class of solutions precise,
suppose that the sigma model target manifold  $N$ is a group manifold $G$, $N=G$,  and $\mathfrak {Lie}G$
is  $\bR^{5,1}$, $\mathfrak{sl}(2, \bR)\oplus \mathfrak{su}(2)$ or $\mathfrak{cw}_6$.
Moreover, we gauge the right action of $G$ on $G$.
The vector fields which generate this action are the left invariant vector fields and $c_{ab}=-\mt_{ab}$.

It remains to explain the relation between the connections $\lambda$ and $C$, and the role of the base space $B$
of $M=P(G, B;\pi)$.
The connection $C$ that we have introduced to gauge the sigma model should not be thought of as
a worldvolume field. Instead, it should be considered  as a composite field which depend
of the scalars of the base manifold $B$. Now the base manifold has also a Wess-Zumino term
$\tilde H$ which is invariant, ${\cal L}_{\xi^a}\tilde H=0$,  and $i_{\xi^a}\tilde H=0$, so
$\tilde H$ has trivial gauging. Now taking $H=W+\pi^*\tilde H$, where $W$ is given in (\ref{gfwz}),  and evaluating $dH$, we
recover   (\ref{closeH}). Furthermore, the minimal coupling of the kinetic term of sigma model action (\ref{sigmact}),
the identification of $N=G$, and the addition of the metric of the base space $B$, which has trivial gauging, reproduce
the metric of the spacetime of the ${\rm hol}(\hat\nabla)\subseteq SU(2)$ class of solutions.

\subsection{Conserved currents and worldvolume supersymmetry}

\subsubsection{(1,1) supersymmetry}

Suppose that the gauge connection $A$ is identified with $\check \nabla$, and so $dH=0$.
In such a case,
the associated worldvolume theory of all generic backgrounds with ${\rm hol}(\hat\nabla)\subseteq SU(2)$
 has (1,1)  supersymmetry. Two-dimensional
supersymmetric sigma models have been extensively investigated  before \cite{gates, sierra, hullwitten, howepap1}.
Here we follow \cite{howepap1}. To describe the worldvolume theory
introduce the worldvolume coordinates $(\sigma^\pp, \sigma^=, \theta^+, \theta^-)$, where
$(\sigma^\pp, \sigma^=)$ are even lightcone  and $(\theta^+, \theta^-)$ are odd coordinates, respectively, and the usual
odd
superspace derivatives
$D_-$ and $D_+$,
$D_-^2=i\partial_=$, $D_+^2=i\partial_\pp$.
Let $g$ be the spacetime metric of the ${\rm hol}(\hat\nabla)\subseteq SU(2)$ backgrounds.
Since $dH=0$,  $H=db$, the worldvolume action written in terms
of (1,1) superfields is
\bea
S=\int d^2\sigma d\theta^+ d\theta^-(g+b)_{MN} D_+X^M D_-X^N~,
\eea
where the spacetime coordinates $x^M$ have been promoted to worldvolume superfields
$X^M(\sigma^\pp, \sigma^=, \theta^+, \theta^-)$. The components of the superfield $X$ are
\bea
X^M=X^M\vert~,~~~\psi^M_-=D_-X^M\vert~,~~~\psi^M_+=D_+X^M\vert~.
\eea

The field equations are
\bea
\hat\nabla_-D_+X^M=0~.
\eea
It is easy to see, that in addition to the energy momentum tensor, and the (1,0) and (0,1)
supersymmetry currents, for every $\hat\nabla$-parallel form there is a conserved current. In particular,
the ``elementary'' conserved currents of generic backgrounds are
\bea
&&T_{++}=g_{MN} D_+X^M D_+ X^N~,~~~T_{--}=g_{MN} D_-X^M D_-X^N~,
\cr
&&J^a=\l^a_M D_+X^M~,~~~U^r=\omega^r_{MN} D_+X^M D_+X^N~.
\la{cur}
\eea
Non-generic backgrounds may have additional currents. As we shall mention later these may lead
to the integrability
of the string dynamics.

\subsubsection{Ultraviolet finiteness}

One of the issues that arises in the worldvolume theory is whether the associated sigma model is ultraviolet finite.
Since the worldvolume theory of generic backgrounds has manifest (1,1) supersymmetry, it is not sufficient
to guarantee ultraviolet finiteness. For this, the worldvolume supersymmetry has to be enhanced
to  (4,1) \cite{howepap1}. In turn (4,1) supersymmetry requires  the spacetime  admits a hyper-complex structure.

As we have already explained the spacetime $M$ of generic ${\rm hol}(\hat\nabla)\subseteq SU(2)$ backgrounds
 admits a hyper-complex type of structure, $I_r$, which is $\hat\nabla$-parallel. However, the endomorhisms
$I_r$ are {\it degenerate} and so cannot be thought of as complex structures on $M$. Moreover since
$M$ is Lorentzian, it does not admit complex structures which are hermitian with the respect the Lorentzian metric.
Nevertheless, we shall provide an indirect argument that the worldvolume theory of some of our
backgrounds  are  ultraviolet finite. First,
all these models are finite up to two loops. This can be shown by  direct observation using the identification
of the $\check\nabla$ connection with the gauge connection. Next we restrict our analysis to the wrapped on $AdS_3\times S^3$
5-brane solutions.
First observe that $M=P(SL(2,\bR)\times SU(2), B_{\rm hk};\pi)=SL(2,\bR)\times P(SU(2), B_{\rm hk};\pi)$. The fields
along $SL(2,\bR)$ decouple from the rest of the theory. So the question is whether the sigma model
on  $P(SU(2), B_{\rm hk};\pi)$  is ultraviolet finite.

To find whether the sigma model on $P(SU(2), B_{\rm hk};\pi)$ is ultraviolet finite consider instead
the sigma model on $P(SU(2)\times U(1), B_{\rm hk};\pi)=P(SU(2), B_{\rm hk};\pi)\times U(1)$, ie add
an additional free superfield $X$ in the sigma model on $P(SU(2), B_{\rm hk};\pi)$. It turns out now that
$P(SU(2)\times U(1), B_{\rm hk};\pi)$ admits a $\hat\nabla$-parallel hyper-complex structure given by
\bea
\Omega^r=\zeta^r_{a'b'} \lambda^{a'} \wedge \lambda^{b'} +\omega^r~,~~~
\eea
where $\zeta^r$ is the hypercomplex structure on $U(1)\times SU(2)=S^1\times S^3$. It is well known
that $S^1\times S^3$ is an HKT manifold. Consequently the sigma model on $P(SU(2)\times U(1), B_{\rm hk};\pi)$
has (4,1) world volume supersymmetry and so it is ultraviolet finite as a consequence of the results
of \cite{howepap1}. Since the $P(SU(2), B_{\rm hk};\pi)$
differs from $P(SU(2)\times U(1), B_{\rm hk};\pi)$ up to the addition of a free superfield, which does not affect the
interactions, the sigma model on $P(SU(2), B_{\rm hk};\pi)$ is also ultraviolet finite. Thus
all wrapped 5-brane solutions on $AdS_3\times S^3$ are ultraviolet finite. We do not have an argument which
generalizes to the rest of solutions. However, it is likely that
all backgrounds of the class ${\rm hol}(\hat\nabla)\subseteq SU(2)$ which are associated with
a (1,1)-supersymmetric worldvolume theory are ultraviolet finite.

\subsection{String integrability}

As we have argued a large class of the supersymmetric heterotic backgrounds with 8 supersymmetries, and possibly  all,
are ultraviolet finite. So, the question that arises is whether string theory can be solved on such backgrounds.
With this we mean to solve the classical embedding equation of a string in such backgrounds, quantize the theory and
construct the Hilbert space of string states. This does not appear to be the case for all backgrounds. For example,
string theory has not been solved on the standard 5-brane background of \cite{callan}. Generic background
do not appear to admit sufficient conserved charges (\ref{cur}) that may lead to integrability. Indeed
 although the theory has a Kac-Moody algebra
as a symmetry, it does not have two copies of it as in the case of other integrable models like the
WZW ones. Nevertheless, string theory is integrable on some special backgrounds which in some cases
appear as limits of generic ones. These special backgrounds are
\bea
G\times B_{hk}~,~~~G\times S^3\times \bR~;~~~B_{hk}=\bR^4~,~~~G=\bR^{5,1}~,~~~AdS_3\times S^3~,~~~CW_6~,
\eea
up to discrete identifications. One can take $B_{hk}$ to be another hyper-K\"ahler manifold instead of $\bR^4$,
or its discrete identification $T^4$, but this choice will suffice in what follows.

All backgrounds $G\times \bR^4$, $B_{hk}=\bR^4$, are either the 10-dimensional Minkowski space or a product
of a WZW model with flat space. Moreover the dilaton is constant. The solution of string theory on
such backgrounds follows from that
on flat spaces and WZW models, see eg \cite{gepner}.
The solution of string theory in Minkowski space is well known.
For the $CW_6\times \bR^4$ background, the solution of
string theory on the plane wave $CW_6$ with
a non-trivial H-flux has been given in \cite{blaugp}. It remains to investigate the solution of string theory
in the $AdS_3\times S^3\times \bR^4$
background. The spectrum of string theory on $AdS_3$ has been investigated in detail in \cite{ooguri}
and references within. Moreover the string spectrum on $SU(2)$ is given by that of the standard WZW model.
The central charge of the bosonic sector of the background  is
\bea
c={3k\over k-2}+{3k\over k+2}+4
\eea
where each term corresponds to the contribution of the associated space in the product $AdS_3\times S^3\times \bR^4$.

The backgrounds $G\times S^3\times \bR$ appear as near horizon geometries of 5-branes with worldvolume $G$.
These are linear dilaton backgrounds. The dilaton depends on the coordinate of $\bR$.
  The solution of string theory on $G$ has already been explained in the $G\times\bR^4$
case above. In addition, the solution of string theory on $S^3\times \bR$ has already been investigated in \cite{callan}
in the context of 5-branes with worldvolume $\bR^{5,1}$. The theory on $S^3$ is a WZW model while
the theory along $\bR$ is a Feigin-Fuks theory associated with the linear dilaton.
 For example the total central charge of the bosonic sector of the
$AdS_3\times S^3\times S^3\times \bR$ background is
\bea
c={3k\over k-2}+{3k\over k+2}+ {3p\over p+2}+ (1+{6\over p})~,
\eea
where $p$ is the 5-brane charge computed at infinity.
The first two contributions to the central charge are due to $AdS_3\times S^3$ worldvolume geometry.  The first of the
other two contributions to the central charge is due to the WZW model associated with the near horizon 3-sphere
while the last contribution
is due to the to a Feigin-Fuks theory.
Unlike the $G\times \bR^4$ case above, the presence of a linear dilaton in these backgrounds gives rise
to regions in spacetime which are strongly coupled and so string loop corrections should be taken into account.

\subsubsection{(1,0) supersymmetry}

So far, we have investigated the worldvolume theory of ${\rm hol}(\hat\nabla)\subseteq SU(2)$ backgrounds
for which the $\check\nabla$ connection has been identified with the gauge one, and so $dH=0$.
If this is not longer the case, then Bianchi identity of $H$ may receive an anomalous
contribution  (\ref{anb}). The worldvolume theory  of generic such backgrounds admits
(1,0) supersymmetry. Let $g$ be the spacetime metric. At the zeroth order in $\alpha'$, $dH=0+{\cal O}(\alpha')$,
and so $H=db+{\cal O}(\alpha')$. In addition let $(\sigma^\pp, \sigma^=, \theta^+)$ be the worldvolume
superspace coordinates, where  $(\sigma^\pp, \sigma^=)$ are commuting lightcone coordinates and
$\theta^+$ is an odd coordinate. Moreover, introduce the superspace coordinate $D_+$,  $D_+^2=i\partial_\pp$. Then the
worldvolume action of string propagating in such backgrounds is
\bea
S=-i\int d^2\sigma
d\theta^+ [(g+b)_{MN} D_+X^M \partial_=X^N+ i h_{ab} \psi_-^a \nabla_+ \psi_-^b]~,
\la{10act}
\eea
where the spacetime coordinates $x^M$ have been promoted to superfields $X^M(\sigma^\pp, \sigma^=, \theta^+)$,
$A$ is the gauge connection, and  $\nabla_+\psi^a_-=D_+\psi_-^a+ DX^M A_M{}^a{}_b \psi^b_-$.
The theory has an additional fermionic superfield $\psi_-$. The components
of both type of superfields are
\bea
X^M=X^M\vert~,~~~\lambda^M_+=D_+X^M\vert~,~~~\psi^a_-=\psi_-^a\vert~,~~~\ell^a=\nabla_+\psi_-^a\vert~,
\eea
where $\ell$ is an auxiliary field.

We have not constructed explicit solutions of this class of backgrounds. However, it is straightforward to see
that one can easily generalize the wrapped 5-brane solutions on $G$ to this case. We shall not give the details here.
We shall focus instead on the issue of ultraviolet finiteness for these models. In particular consider
the case of wrapped 5-brane solutions on $AdS_3\times S^3$ for which $S^3$ is gauged but the
$\check\nabla$ connection is not identified with the gauge one. Generalizing  the argument we have presented
for the backgrounds with (1,1) worldvolume supersymmetry,  we can show that such (1,0)-supersymmetric backgrounds
are also ultraviolet finite.
However in this case, the proof for ultraviolet finiteness is based on that of (4,0)-supersymmetric
sigma models \cite{howepap1, howepap2} instead of that of (4,1) supersymmetric ones. It is known
that the couplings of (4,0)-supersymmetric sigma models receive corrections from changing the renormalization
scheme from a (1,0)-supersymmetric  to manifestly (4,0)-supersymmetric one \cite{howepap2}.
Such corrections are consistent with the contribution from the curvature square terms in (\ref{fineqn}).

\newsection{Holonomy $\bR^8$ solutions}

\subsection{Geometry}

The backgrounds with 8 supersymmetries and ${\rm hol}(\hat\nabla)\subseteq \bR^8$ \cite{het1} admit a
light-cone orthonormal frame $e^A=(e^-, e^+, e^i)$,
 $i=1,6,2,7,3,8,4,9$. In this frame, the conditions that arise  from the Killing spinor equations can be written as
\bea
&&{\rm hol}(\hat\nabla)\subseteq \bR^8~,~~~\hat\nabla e^-=0~,~~~\partial_+\Phi=0~,~~~de^-\in \bR^8~,~~~
\cr
&&H_{ijk}=0~,~~~2\partial_i\Phi-H_{-+i}=0~.
\la{r8con}
\eea
Moreover, the spacetime metric and $H$ are
\bea
ds^2&=&2 e^- e^++ \delta_{ij} e^i e^j~,
\cr
H&=&e^+\wedge de^-+\Omega_{-,ij} e^-\wedge e^i\wedge e^j~,
\eea
where $\Omega$ is the spin connection.
To solve the conditions that arise from the Killing spinor equations, first observe that
(\ref{r8con}) implies that the vector field $X$  associated to the null 1-form $e^-$ is Killing.
Adapting a coordinate to $X$ as $X=\partial_u$. The metric is independent of $u$ since $X$ is an isometry.
Next the condition $de^-\in \bR^8$ can also be written as
\bea
e^-\wedge de^-=0~.
\eea
This implies that there is a coordinate $v$ such that
\bea
e^-=h^{-1}(y, v) dv~,
\eea
for some function $h$ which depends on the coordinates $(v, y^I)$, where $y^I$ are the coordinates along the
transverse directions to the light cone.

In terms of the coordinates $(u,v, y^I)$, the  frame can be written as
\bea
e^-= h^{-1}(y,v) dv~,~~~e^+=du+V dv+n_I dy^I~,~~~e^i= e^i_I dy^I+ t^i dv~.
\eea
This frame is not unique. There is a residuum $\bR^8$-gauge symmetry which leaves $e^-$ and the Killing spinors invariant
which can be used to set $t=0$. So one has
\bea
e^-=h^{-1} dv~,~~~e^i= e^i_I dy^I~,~~~e^+=du+V dv+n_I dy^I~,
\eea
where $h, e^i_I, V$ and $n_I$ depend on both $v$ and $y^I$ coordinates.
The metric and 3-form field strength can be written as
\bea
ds^2&=&2 h^{-1} dv (du+V dv+n_I dy^I)+\delta_{ij} e^i_I e^j_J dy^I dy^J
\cr
H&=&e^+\wedge d e^--\big(h\, \delta_{k[i} e^J_{j]}\,\partial_v e^k_J+\partial_{[I} n_{J]} e^I_i e^J_j\big)\, e^-\wedge
e^i\wedge e^j
\eea

Using the first condition in (\ref{r8con}), $H_{ijk}=0$,
 and the torsion free conditions for the frame $(e^-, e^+, e^i)$, one finds that
\bea
&&\partial_I(2\Phi+\log h)=0~,~~~\partial_{[I} e^i_{J]}=0~,~~~\Omega_{-,-i}= e^I_i (-\partial_I V+\partial_v n_I)~,
\cr
&&\Omega_{-,ij}=-h\, \delta_{k[i} e^J_{j]}\,\partial_v e^k_J-\partial_{[I} n_{J]} e^I_i e^J_i~.
\la{torfree}
\eea
The first condition relates the dilaton to $h$. The second condition implies that there are functions $x^i(y^J, v)$ such that
\bea
e^i_I=\partial_I x^i~.
\eea
Performing the coordinate transformation $x^i=x^i(y^I, v)$ and after an appropriate redefinition of $V$ and $n$,
the fields can be rewritten as
\bea
&&ds^2=2e^- e^++\delta_{ij} dx^i dx^j~,~~~~~~e^-=h^{-1} dv~,~~~e^+=du+  V dv + n_i dx^i~,
\cr
&&H=d (e^-\wedge e^+)~,~~~e^{2\Phi}= h^{-1} g(v)~,
\la{solksea}
\eea
where $h,  V,  n$ are functions of $v$ and $x^i$, and $g$ is an arbitrary function of $v$.
This concludes the analysis of the gravitino and dilatino Killing spinor equations.

The gaugino Killing spinor equation implies
\bea
F_{+i}=F_{ij}=F_{-+}=0~,
\eea
where we have suppressed the gauge indices.
In the coordinates of (\ref{solksea}), these can be written as
\bea
F_{ui}=F_{ij}=F_{vu}=0
\eea
Choosing the gauge $A_u=0$, we have that the above conditions imply
\bea
A_i=A_i(v, y)~,~~A_v=A_v(v, x^i)~,~~~A_i=U^{-1}\partial_i U~.
\eea
After performing a gauge transformation using $U$, the only non-vanishing component of $A$ is $A_v$ and
\bea
F_{vi}=\partial_i A_v~.
\la{solkseb}
\eea
Clearly
the one-loop contribution  to the Bianchi identity of $H$ due to the anomaly cancelation mechanism
for both $A$ and $\check \nabla$ connections vanishes. So the
solutions are at least two-loop exact.

It is clear from the analysis so far that to find explicit solutions in this case, one has to determine the
unknown functions that appear in (\ref{solksea}) and (\ref{solkseb}). These are $h(x,v), V(x,v), n_i(x,v)$ and $A_v(x,v)$.
This is done by solving the field equations.

\subsection{The solution of the field equations}

The Killing spinor equations imply some of the field equations.
It is known \cite{het1, het2} that the remaining independent field equations that have to be solved
are   the   $E_{--}=0$  component of the Einstein equations, the field equations $LH_{AB}=0$
of the 2-form gauge potential and the $LF_-=0$ component of the 1-form gauge potential.
The field equation $LH_{AB}=0$ gives
\bea
\partial_i^2h=0~,~~~\partial^j(dn)_{ji}+g \partial_v(g^{-1} \partial_i h)=0~.
\la{sol1}
\eea

Using $\partial_i^2h=0$, the $E_{--}=0$ component of the Einstein  equations implies that
\bea
&&h^{-1} \partial_i^2 V- h^{-1} \partial_v\partial^in_i-\partial_v^2 \log(h^{-1} g)+ h\partial_vh^{-1}
\partial_v \log(h^{-1} g)=0~.
\la{sol2}
\eea

Moreover, $LF_-=0$ gives
\bea
\partial_i^2 A_v=0~.
\eea
and so $A_v$ is a harmonic function on $\bR^8$.

The above equations of motion can be simplified somewhat provided we take $g=1$ and choose
\bea
\partial^i n_i=\partial_v h~.
\la{ccc}
\eea
In such a case, they can be written as
\bea
\partial_i^2 h=0~,~~~\partial_i^2 n_j=0~,~~~\partial_i^2V=0~,~~~\partial_i^2 A_v=0~.~~~
\eea
Therefore $h$, $n$, $V$ and $A_v$ are $v$-dependent harmonic functions of $\bR^8$. Moreover, it is required
that $n$ and $h$ satisfy (\ref{ccc}).
Solutions of these equations that depend on both $v$ and $x$ coordinates have been given in \cite{kostas}.

Alternatively, one can assume that $h, V, n$ and $A_v$ depend only on $x$ and $g=1$. In such a case, one finds that
\bea
\partial_i^2h=0~,~~~\partial^j(dn)_{ji}=0~,~~~\partial_i^2 V=0~,~~~\partial_i^2 A_v=0
\eea
So again $h, V$ and $A_v$ are harmonic functions of $\bR^8$ and $n$ satisfies the Maxwell equations on $\bR^8$.
It is  clear that  the most general solution
of this class is the superposition of a fundamental string with a pp-wave and with a null rotation.

\subsection{Worldvolume theory}

As we have already mentioned there is no anomalous contribution to the Bianchi identity of $H$
in the ${\rm hol}(\hat\nabla)\subseteq \bR^8$ class of backgrounds, and so $dH=0$. However,
the gauge connection $A$ need not be identified with $\check \nabla$. In fact in the context
of the heterotic string $A$ cannot be identified with $\check \nabla$ because the holonomy of the  former is compact
while the holonomy of the latter
is Lorentzian for generic backgrounds \cite{tseytlin1}.  As a result the worldvolume
theory of generic backgrounds  is (1,0)-supersymmetric. The action is given as in (\ref{10act})
and we shall not repeat the analysis.

To describe the ultraviolet properties of ${\rm hol}(\hat\nabla)\subseteq \bR^8$ backgrounds we shall consider two cases.
If the solutions are taken as backgrounds of the common sector of type II strings, ie the gauge
connection is identified with $\check\nabla$, and so the the worldvolume supersymmetry is enhanced to (1,1), then
it has been shown in \cite{tseytlin1} that  worldvolume theory is ultraviolet finite provided $\partial_vh=0$.
Next consider the ${\rm hol}(\hat\nabla)\subseteq \bR^8$ backgrounds as solutions of the heterotic string. As it has
already
been mentioned
${\rm hol}(\check\nabla)$ cannot be identified with $A$ for generic backgrounds. So in general the worldvolume theory
has strictly (1,0) worldvolume supersymmetry. In addition it is known that in many cases, the 2-loop beta function
does not vanish. An exhaustive analysis of the ultraviolet properties of these backgrounds in the context of
heterotic string can be found in \cite{tseytlin1}.

\newsection{Conclusions}

We have described all backgrounds of heterotic strings  that preserve half the supersymmetry, ie 8 supercharges.
There are three classes of backgrounds distinguished by the holonomy of the connection with torsion $\hat\nabla$.
The backgrounds of the ${\rm hol}(\hat\nabla)\subseteq SU(2)$ class can be constructed starting from any
4-dimensional hyper-K\"ahler manifold, $B_{hk}$, and a anti-self-dual connection on it with gauge group $G$ such
that $\mathfrak{ Lie}(G)=\bR^{5,1}, \mathfrak{sl}(2,\bR)\oplus\mathfrak{su}(2), \mathfrak{cw}_6$ is a self-dual
Lorentzian Lie algebra. In particular, the spacetime is a principal bundle with base space $B_{hk}$, fibre group
$G$ and equipped with a compatible anti-self-dual connection $\lambda$.
We demonstrate that the generic solutions of this type have the interpretation
of wrapped 5-branes on $G$.

We have constructed new solutions which are characterized by 3 integers and
depend on a continuous moduli which includes  that of $SU(2)$ instantons. There
may be a relation of some of these backgrounds to the ADHM sigma models of \cite{witten}. We also
explore their interpretation in M-theory in terms of M2- and M5-brane intersections.

We have investigated the worldvolume theory of ${\rm hol}(\hat\nabla)\subseteq SU(2)$ backgrounds and show that
it is closely related to that of gauged WZW models. However the worldvolume gauge field in this case
is not independent but rather it is  a composite that depends on the scalars associated with $B_{hk}$.
We demonstrate that a large class of the worldvolume theories based on these backgrounds are ultraviolet finite. This
may be extend  to all ${\rm hol}(\hat\nabla)\subseteq SU(2)$ backgrounds.

We also show that the backgrounds of the ${\rm hol}(\hat\nabla)\subseteq \bR^8$ class are superpositions
of the fundamental string with a pp-wave which may also include a null rotation.

{}For the third class of backgrounds that preserve 8 supersymmetries, ${\rm hol}(\hat\nabla)=\{1\}$. The spacetime
is a Lorentzian group manifold. These backgrounds have been classified in \cite{josek}.

The analysis we have presented is complete for the ${\rm hol}(\hat\nabla)\subseteq \bR^8$ class of backgrounds.
The description of all ${\rm hol}(\hat\nabla)\subseteq SU(2)$  solutions requires the classification of all 4-dimensional
hyper-K\"ahler manifolds and their  anti-self-dual instantons which have gauge group $G$ with Lie algebra
$\mathfrak{sl}(2,\bR)\oplus\mathfrak{su}(2)$ or $\mathfrak{cw}_6$. There are explicit constructions of these data in many special
cases but the  description is not complete.
Even in the case that $B_{hk}=\bR^4$  to construct all solutions, one has to find the (non-abelian) instantons
with a Lorentzian gauge group $G$;
$\mathfrak{Lie} G=\mathfrak{sl}(2,\bR)\oplus\mathfrak{su}(2), \mathfrak{cw}_6$.
Unlike for instantons on $\bR^4$ with a Euclidean gauge group, little is known about  Lorentzian gauge group case. Though related
configurations  have been investigated in eg \cite{mosna}.
Since any 4-dimensional hyper-K\"ahler manifold can be used  to construct solutions,
one may consider other examples from that with $B_{hk}=\bR^4$. An alternative  choice is $B_{hk}=T^4$
and the anti-self-dual connections on $T^4$ to give new solutions. The supersymmetry of such
backgrounds will depend on the choice of spin structures on $T^4$ and this worths further investigation.
New explicit solutions can also be given by taking $B_{hk}$ to be the Gibbons-Hawking manifold.

It is clear that there are continuous families of ${\rm hol}(\hat\nabla)\subseteq SU(2)$ backgrounds
and their moduli is very large. Nevertheless all such backgrounds can be  understood in terms of
4-dimensional hyper-K\"ahler geometry and the theory of instantons on hyper-K\"ahler manifolds.

\vskip 0.5cm
{\bf Acknowledgments:}~I would like to thank Ulf Gran  and Arkady Tseytlin for helpful discussions.

\end{document}